# Light Programmable Micro/Nanomotors with Optically Tunable In-Phase Electric Polarization

**Zexi Liang[†], Daniel Teal[‡], Donglei Fan[†‡][*]**

[†] Materials Science and Engineering Program, The University of Texas at Austin, Austin, TX, 78712

[‡] Department of Mechanical Engineering, The University of Texas at Austin, Austin, TX, 78712

[*] Corresponding author. Email: dfan@austin.utexas.edu

## Abstract:

To develop active nanomaterials that can instantly respond to external stimuli with designed mechanical motions is an important step towards the realization of nanomachines and nanorobots. Herein, we present our finding of a versatile working mechanism that allows instantaneous change of alignment direction and speed of semiconductor nanowires in an external electric field with simple visible-light exposure. The light induced alignment switch can be cycled over hundreds of times and programmed to express words in Morse code. With theoretical analysis and numerical simulation, the working principle can be attributed to the optically tuned real-part (in-phase) electrical polarization of a semiconductor nanowire in an aqueous suspension. The manipulation principle is exploited to create a new type of microscale stepper motor that can readily switch between in-phase and out-phase modes, and agilely operate independent of neighboring motors with patterned light. This work could inspire the development of a new type of micro/nanomachines with individual and reconfigurable maneuverability for many applications.

## Introduction:

The future micro/nanorobots require a high degree of freedom in motion control to perform complex tasks by individuals or by a swarm. However, it is a daunting task to apply the established technologies that manipulate macroscopic robots to realize similar functions in their micro/nanoscale counterpart. There are several grand challenges, ranging from designing and fabrication of active nanocomponents, precision device assembly, to actuation with desired performances[1]. Furthermore, physics that governs mechanical motions at micro/nanoscales is distinctive from that over the large scale. In addition to the level of required force to operate a micro/nanoobject being reduced by orders of magnitude, non-important forces in macroscopic regime, such as electrostatic interactions, become prominent in micro/nanoscale. Furthermore, since most micro/nanomachines operate in suspensions, the suspension related low Reynolds number physics[2], ionic effects[3], and interactions of suspension medium with energy sources emerge for consideration[4]. To address these challenges, in the past decades, rapid progresses have been made in basic science[5,6], fabrication approaches[7], assembling strategies[8-15], and a variety of powering mechanisms of micro/nanomachines[16-21]. An array of potential applications have been demonstrated in biosensing, cargo and molecular delivery, single cell bioresearch, and environmental remedy[22-26].

However, it remains a great challenge to control the motions of an individual nanomachine amidst many, switch the operation modes facilely, and it is even more difficult to actuate several components of a nanomachine coordinately for purposed actions[27]. This high degree of versatility is essential for the future micro/nanorobots and requires investigation of innovative actuation mechanisms. Recently, research interest has been focused on utilizing two different propulsion mechanisms to generate stimulus responsive motions of nanomachines that can alter among different types of motions[28]. For instance, catalytic and optochemical nanomotors can instantly align, stop, and even change propulsion directions in an external magnetic[29], or electric field[30]. However, these demonstrations are largely based on combining two well understood propulsion mechanisms to achieve switchable motion control.

It has been a rarely exploited concept to generate switchable operations of nanomotors among different modes via instantaneously tuning their physical properties, *e.g.* electric conductivity[31] with an external stimulus. Recently, our finding shows that visible light is able to change the imaginary part of electric polarization of semiconductor nanowires, such as silicon, as reflected by the dramatic change of rotation orientation and speed in a high-frequency rotating electric field[31]. According to the Kramers-Kronig relation, the imaginary part and real part of electric polarizations are correlated[32]. It suggests that if the imaginary part of electric polarization alters with an external stimulus, the correlated real part will also respond to the same stimulus, which however, has not been experimentally verified in this scenario.

Mechanical alignment of a nanowire by an electric field is a result of the interaction between the in-phase (real part) electric polarization of the nanowire and the electric field. Here, we report our study of the light effect on the electric alignment of semiconductor nanowires made of Si. With visible light, a nanowire can be readily toggled between different alignment speeds and orientations. The switchable manipulation is instant, robust, and programmable, which operates for hundreds of cycles and can be programmed to communicate words in Morse code. Theoretical analysis and numerical simulation are carried out to unveil the underlying physical mechanism, which is attributed to the light tunable in-phase (or real part) electric polarization of semiconductor nanoparticles in suspension. The results agree with experiments qualitatively and the inference from the Kramer-Kronig relation. With the obtained understanding, we propose and successfully realize the first light switchable stepper micromotor based on the tunable in-phase electric polarization. With designed light patterns projected by a digital light processing system (DLP), nanomotors demonstrate independent mechanical operations amidst neighbors in the same electric field.

Silicon nanowires are fabricated via the well-known metal-assisted chemical etching (MACE) methods with details described in the methods section. All the nanowires studied in our experiments are approximately 10 μm in length and 500 nm in diameter, made from undoped silicon as shown in the SEM image in Fig. 1A and B. Two types of microelectrodes made of Cr (5 nm)/Au (100 nm) films are designed and lithographed on glass substrates as shown in Fig. 1C. The parallel microelectrodes are made of two rectangular metal pads separated by a uniform gap of 125 μm (Fig. 1D), via which a spatially uniform electric field can be generated in the center of the gap for measurement of the alignment rate of silicon nanowires from 5 kHz to 4 MHz and up to 15 $V_{pp}$. The quadruple microelectrodes are made of four trapezoidal electrode pads, patterned orthogonally, creating a square central area of 500 μm in side length (Fig. 1E). The quadruple microelectrodes are utilized to realize and characterize the light controlled microscale stepper motor to be discussed later.

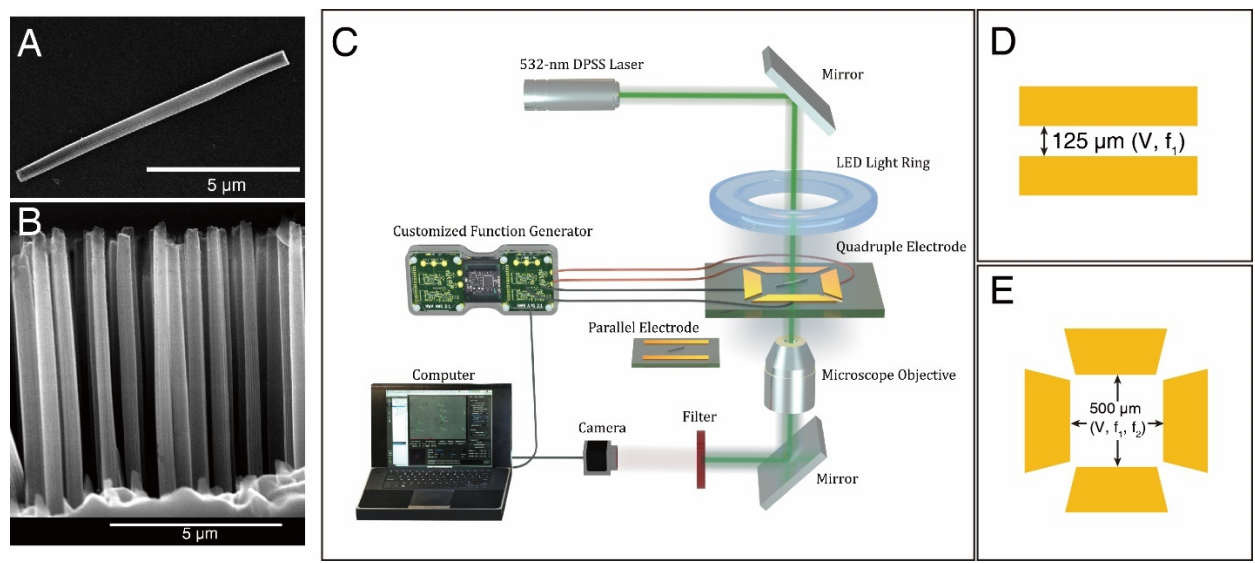

Figure 1. (A, B) SEM images of the fabricated undoped silicon nanowires with 10-μm length and 500-nm diameter. (A) A single silicon nanowire. (B) Cross section image of silicon nanowire arrays. Schematics of (C) experimental setup for switchable alignment of nanowires and rotation phase of microscale stepper motors; (D) parallel microelectrodes for electro-alignment; (E) and quadruple microelectrodes for actuation of the rotary stepper motors.

## Results:

**Optically tunable alignment of silicon nanowire in AC electric field.**

When placed in an AC electric field in the range of kHz to MHz, silicon nanowires can be readily aligned in response to the field. This effect has been observed on longitudinal structures and is known as electro-alignment[33]. However, different from previous works, for Si nanowires, which normally align parallel to the applied electric field across a broad range of AC frequencies, we also observe perpendicular alignment to the electric field in a narrow range of frequencies in deionized water. Moreover, when illuminated with laser (532 nm), the alignment torques on silicon nanowires significant change, resulting in acceleration of alignment speed and even switch between parallel and perpendicular alignment directions, which can be cycled for hundreds of times continuously. This switchable manipulation has been observed for the first time. To understand this effect, we carry out a series of experiments, characterize the behaviors, and conduct theoretical analysis with numerical simulation.

It is known that when there is no external electric field, a nanowire suspended in water exhibits Brownian motions and is randomly oriented. For instance, at time $t_1$, a nanowire orients at $\theta_1$ (θ is the angle between the long axis of the nanowire and the electric field). As soon an electric field is applied via the parallel microelectrodes, the nanowire rotates to align in response to the electric field. At time $t_2$, the angle is $\theta_2$. The alignment rate (A) can be readily determined as shown in supplementary note 2:

$$A = -\frac{1}{t_2 - t_1} \int_{\theta_1}^{\theta_2} \frac{d\theta}{\sin\theta\cos\theta} \quad (1)$$

Most frequently, the nanowire aligns parallel to the electric field with $\theta_2 \approx 0$, which is termed as parallel alignment with a positive alignment rate. Interestingly, the nanowire also aligns perpendicular to the electric field with $\theta_2 = \pi/2$ under certain conditions, which is termed as transverse alignment with a negative alignment rate.

By sweeping the AC frequency and toggling the laser on and off, we obtain the alignment behaviors of Si nanowires at different frequencies with (red curve) and without laser (blue curve) in Fig. 2A. Here, a minimum LED illumination (white light ~500 lux) is used as background lighting to record motions of nanowires when there is no laser. A 532-nm laser at 32

mW/cm$^2$ is applied together with the LEDs illumination to obtain the laser induced alignment behaviors in the red curve.

With the minimum background light and no laser exposure, the frequency dependent alignment of a nanowire can be categorized into three regimes: (1) In the low frequency regime from 5 kHz to 500 kHz, nanowires align parallel to the applied electric field with a maximum alignment rate of 150 rad/s at 5 kHz. The rate gradually decreases with AC frequency with a steep drop at around 100 kHz; (2) In between 500 kHz and 1 MHz, the nanowires are aligned perpendicular to the electric field; 3) When the AC frequency further increases to above 1 MHz, the nanowires return to the parallel alignment at a low rate below 10 rad/s. In contrast, when a 532-nm laser is illuminated on the silicon nanowires, dramatic changes of the alignment behaviors, including acceleration and even switching of direction are obtained as shown in the red curve in Fig. 2A. For instance, at 5 kHz, the average alignment rate increases by 2.4 folds to 360 rad/s. The maximum alignment rate increases to 487 rad/s, which is blue shifted to ~25 kHz from that without the laser, and is almost four times that of the same frequency without the laser. As the frequency further increases, the alignment rate decreases rapidly at about 250 kHz and approaches zero at 4 MHz, the highest AC frequency that we test. Overall, laser exposure increases the positive alignment rate at all frequencies, blue-shifts the frequency where the highest alignment rate occurs, and changes the transverse alignment to parallel alignment at the tested frequencies.

To further understand the observed light controlled switching behaviors of Si nanowires, we monotonically vary the laser intensity from 8 mW/cm$^2$ to 127 mW/cm$^2$ and measure the alignment rate as shown in Fig. 2B. For instance, at 5 kHz, the alignment rate increases with laser intensity up to 32 mW/cm$^2$, then maintains an approximate plateau to 64 mW/cm$^2$, before decreasing at 127 mW/cm$^2$. In general, the maximum alignment rate among all these curves increases with laser intensity and the corresponding frequency exhibits a monotonic shift from 5 kHz at 0-16 mW/cm$^2$ to ~100 kHz at 127 mW/cm$^2$. The amplitude of the negative alignment rate decreases with laser power up to 16 mW/cm$^2$, above which the transverse alignment switches to parallel alignment.

**Light programmable toggle of nanowires between two alignment orientations to express words in Morse code.**

Finally, we use this effect as a light-controlled parallel-to-transverse alignment switch. We expose a nanowire in the electric field of 750 kHz and 15 Vpp, and to the 111 mW/cm$^2$ 532-nm laser, which is periodically occluded via a motorized shutter. The nanowire can be switched instantly between parallel and transverse alignment to the electric field, when the laser is on and off, respectively. The switch periodicity follows the on/off state of the laser as shown in Fig. 2C (see supplementary video 2), demonstrating the robustness of the light-controlled switch of semiconductor nanowires in an AC electric field. The high controllability and reproducibility of the light controlled alignment are further exploited by toggling a nanowire to express words, *i.e.* "HELLO WORLD", in Morse code (International Telecommunication Union standard). Here, the duration of the alignment towards and at 90° when the laser is off is used to transmit Morse code, *i.e.* a short signal of dot (·) takes a third of the time of a long signal of dash (-). When the laser is on, the corresponding duration of alignment motion towards and at 0° forms the separation space between signals. In experiments, nanowires indeed switches back and forth rapidly with controlled durations in response to the programmed light signals according to Morse code (Fig. 2D, supplementary video 10, and supplementary file 1). The toggled motions correlate with the light signals well, and can be readily interpreted as "HELLO WORLD" (supplementary file 2). This demonstration is one of the first that utilize mechanical motions of a nanowire to transmit meaningful words, which could be a unique communication approach for the future nanomachines and nanorobots

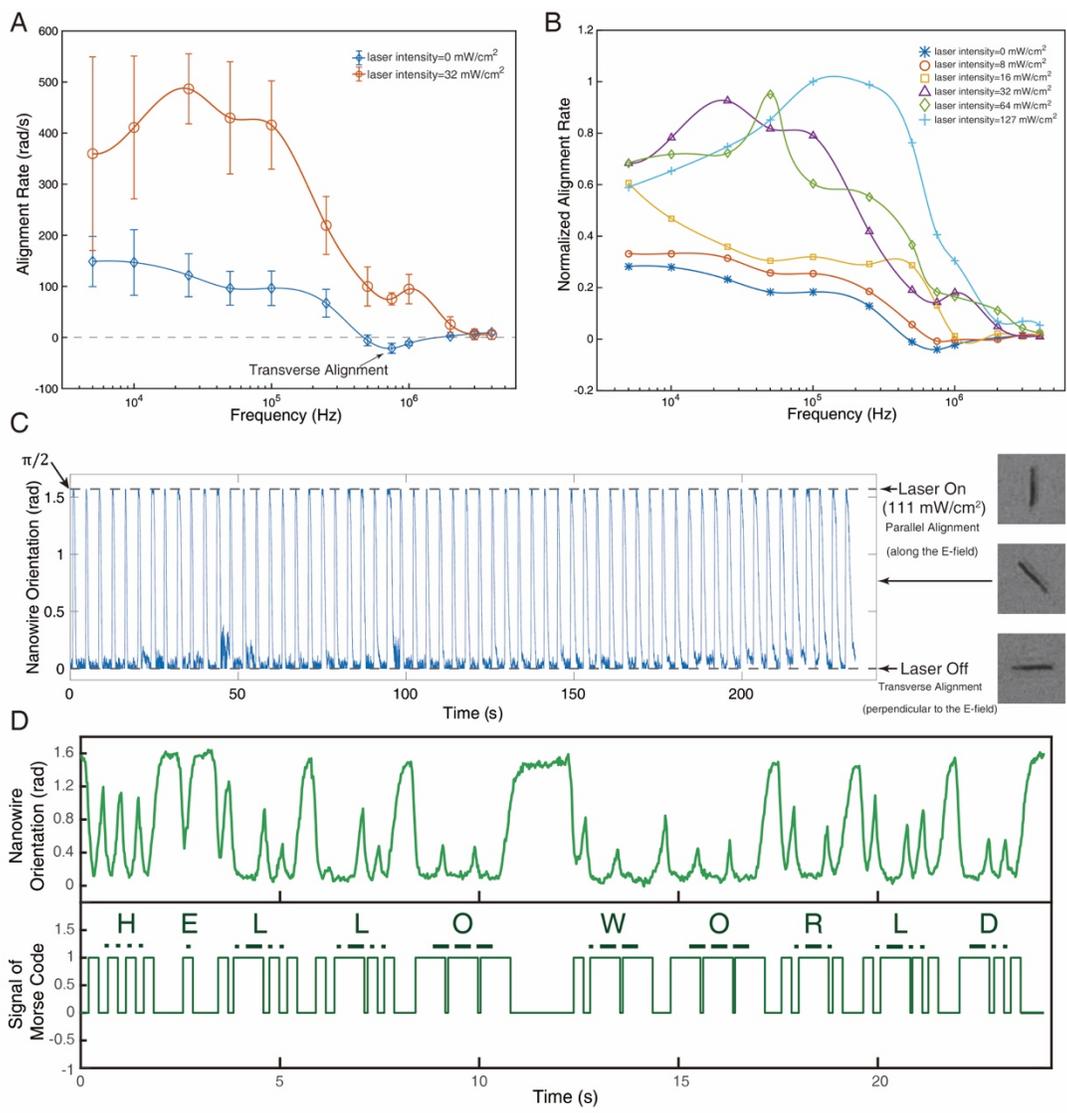

**Figure 2.** Light responsive alignment of Si nanowires. (A) Alignment rate versus AC frequency of silicon nanowires (undoped; length, 10 μm; diameter, 500 nm) in a dim condition (blue) and under 532-nm laser illumination (32 mW cm$^{-2}$, red). (B) Intensity effect of laser on alignment rate versus AC frequency. The eye-guiding lines in (A) and (B) show trends. (C) Cyclic switch between parallel and transverse alignment of a single nanowire with on/off of a 532 nm laser.

The electric field direction is defined as $\pi/2$ and its transverse direction is defined as 0. (D) Toggling of a single nanowire in response to a light signal encoded by Morse code (top) and the corresponding interpreted words "HELLO WORLD" from the mechanical motions of the nanowire (bottom).

**Modeling of the optically tunable in-phase electric polarization of silicon nanowires.**
To understand the above unique light-controlled alignment, we carry out theoretical analysis followed by numerical simulation and modeling. In the presence of a uniform AC electric field, an electric dipole moment ($p$) is induced in a silicon nanowire. When the dipole moment is directed at an arbitrary angle to the electric field, an electric torque, given by $\boldsymbol{\tau} = \boldsymbol{p} \times \boldsymbol{E}$, exerts on the nanowire and orients it until the electrostatic energy is minimized. For simplicity, we consider only a 2D case, where the motions and the involved forces are both in plane with the microelectrodes. We can express the electric field along the x-axis with $\boldsymbol{E} = \mathrm{Re}[E_0\hat{\boldsymbol{x}}\exp(i\omega t)] = \mathrm{Re}[\underline{\boldsymbol{E}}\exp(i\omega t)]$, where $\underline{\boldsymbol{E}} = E_0\hat{\boldsymbol{x}}$ is the phasor of electric field, $E_0$, $\omega$, and $t$ are the amplitude of electric field, angular frequency of the electric field, and time, respectively. The dipole moment ($p$) of the nanowire can be decomposed into two components that are parallel and perpendicular to the long direction of the nanowire, given by $\underline{\boldsymbol{p}_\parallel} = \alpha_\parallel \underline{\boldsymbol{E}_\parallel}$ and $\underline{\boldsymbol{p}_\perp} = \alpha_\perp \underline{\boldsymbol{E}_\perp}$, respectively. Here $\underline{\boldsymbol{p}_i}, \alpha_i, \underline{\boldsymbol{E}_i}$ ($i = \parallel$ or $\perp$) are the phasor of dipole moment, complex polarizability, and the phasor of the electric field along the respective directions. Since the frequency of the AC electric field employed in our experiments is much higher than that of the rotational motion during nanowire alignment, the time-averaged torque ($\tau$) exerted on a nanowire results in the observed alignment, which can be expressed as:

$$\tau_e = \frac{1}{2}\text{Re}\left[\underline{p} \times \underline{E}^*\right] = -\frac{1}{2}E_0^2 \text{Re}(\alpha_\parallel - \alpha_\perp)\sin\theta\cos\theta\,\hat{z} \tag{2}$$

where $\theta$ is the angle between the long axis of the nanowire and the electric field, and $\underline{E}^*$ denotes the complex conjugation of the electric field phasor.

Owing to the low dimensions, a nanowire is in an extremely low Reynolds number regime, where viscous force dominants motion and the drag torque is proportional to rotation speed. The driving electric torque on the nanowire is balanced by the liquid drag torque essentially instantly, given by $\tau_e = -\tau_{drag} = \gamma\dot{\theta}$, where $\gamma$ is the rotational drag coefficient of a nanowire in deionized water. As a result, the rotation speed of a nanowire during an electric alignment depends on the instantaneous angle ($\theta$) between the long direction of the nanowire and the electric field, and is given by:

$$\dot{\theta} = -A\sin\theta\cos\theta \tag{3}$$

where $A = \frac{E_0^2}{2\gamma}\text{Re}(\alpha_\parallel - \alpha_\perp)$ is defined as the alignment rate. Note that equation (3) implies the nanowire rotation speed is zero at angles $\theta = 0$ or $\pi/2$ and half of A when $\theta = \pi/4$. The alignment rate (A) can be experimentally determined by equation (1). Furthermore, since A is linearly proportional to $\text{Re}(\alpha_\parallel - \alpha_\perp)$, one can readily determine the real part of nanowire polarization at a given voltage ($E_0$) and viscous coefficient constant ($\gamma$). A positive alignment rate (A>0) indicates $\text{Re}(\alpha_\parallel - \alpha_\perp) > 0$, where the real part of electric polarization along the parallel direction of a nanowire is greater than that along the transverse direction, $\text{Re}(\alpha_\parallel) > \text{Re}(\alpha_\perp)$. Vice versa, a negative alignment rate (A<0) corresponding to $\text{Re}(\alpha_\parallel - \alpha_\perp) < 0$ or $\text{Re}(\alpha_\parallel) < \text{Re}(\alpha_\perp)$.

To obtain a more qualitative understanding of the observed optically controlled alignment phenomena, we conduct modeling with a method including both analytical and numerical approaches. Related to this work, modeling of the out-of-phase (imaginary part) polarization of semiconductor nanoparticle, which is reflected by the rotation of the particle in a high-frequency rotating AC electric field, has been reported by using either a numerical method[34] or an analytical approach[31]. Different from previous works, here, we focus on understanding optically controlled alignment in an electric field which is governed by the in-phase electric polarization (real part) and leverage the accuracy of numerical simulation and simplicity of analytical approximation to investigate the system.

We consider two major frequency dependent effects in the model, Maxwell-Wagner interfacial polarization and the electrical double layer (EDL) charging. Maxwell-Wagner polarization originates from the difference between electric properties of a particle and the suspension medium, which results in a net electric dipole moment in the particle.[33] With the well-established theory of Maxwell-Wagner polarization, the frequency dependent dipole moment of the nanowire can be calculated analytically. However, in the analytic models of a nanowire, the shape is approximated as a prolate ellipsoid. Here, with numerical simulation based on finite element analysis, we instead can take the accurate wire shape as a cylinder in the model as shown in Fig. 3A, implemented with COMSOL. Furthermore, it is known that as-fabricated Si nanowires inevitably has a native $SiO_2$ layer. Here, we can readily include an oxidation layer of 1-nm on the silicon nanowire in the model to reflect the nature of the system.

In the simulation, the Maxwell-Wagner polarization contributed dipole moment in the nanowire is calculated by directly integrating the electrical potential and first order Legendre polynomials $P_1 = \cos\theta$ along a circular loop encompassing and centered on the nanowire of an azimuthal symmetry about its long axis[35]. There is no such symmetry about the transverse axis, but a sufficiently large circular loop for integration with the same method makes a good approximation. (More details can be found in supplementary note 3.)

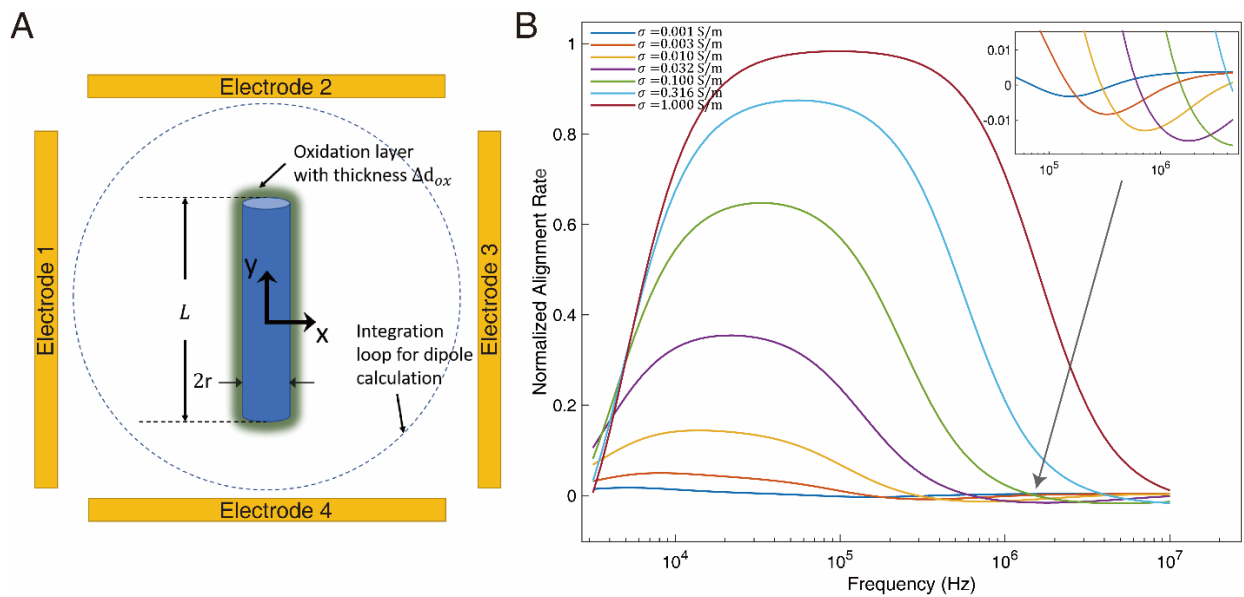

Figure 3. Modeling of nanowire polarization and numerical calculation of alignment rate versus AC frequency of a silicon nanowire of $L$ = 10 μm and $r$ = 250 nm. (A) Model of a silicon nanowire as a cylinder with length $L$ and radius $r$. An Oxidation shell with thickness $\Delta d_{ox} = 1\ nm$ is considered, shown as the dark shade surrounding the nanowire. (B) Numerical calculation of alignment rate of nanowires of different electrical conductivities versus AC frequency. Inset: Zoom-in image at frequencies with negative alignment rates.

Next, we model the electrical double layer around the electrically polarized nanowire. The electric double layer is a result of the Maxwell-Wagner polarization induced surface charge on a nanowire. An equivalent resistor-capacitor (RC) model is used to predict the frequency dependent behavior. Since the EDL is formed by ions in suspension with finite mobilities, there is a phase lag ($\delta$) between the Maxwell-Wagner polarization and EDL charging. At a low-frequency electric field, ions can move quickly enough to make the phase lag $\delta$ small so that the EDL charging can partially or fully screen the surface charges, resulting in an opposite dipole moment to the direction of Maxwell-Wagner polarization. At higher frequencies, the phase lag gradually increases, and eventually, the effect of the EDL vanishes since the ionic diffusion speed cannot catch up the speed of Maxwell-Wagner polarization. We can readily calculate the imaginary and real parts of the EDL charging dipole as [31]:

$$\text{Im}(P_{EDL}) = -\frac{[\text{Re}(P_{MW})\sin\delta + \text{Im}(P_{MW})\cos\delta]}{\sqrt{\omega^2 \tau_{RC}^2 + 1}}$$

$$\text{Re}(P_{EDL}) = -\frac{[\text{Re}(P_{MW})\cos\delta - \text{Im}(P_{MW})\sin\delta]}{\sqrt{\omega^2 \tau_{RC}^2 + 1}}$$

(1)

respectively, where $P_{MW}$ is the dipole moment from Maxwell-Wagner polarization, $\delta$ is the phase lag between polarization and the EDL charging that follows $\tan\delta = -\omega\tau_{RC}$, $\omega$ is the angular frequency of the electric field, and $\tau_{RC}$ is the time constant of the RC model. The total electric dipole moment is the summation of both components of Maxwell-Wagner and the EDL polarizations, given as $P_{total} = P_{MW} + P_{EDL}$.

We simulate the normalized alignment rate to compare with the experimental results. As shown in Fig. 3B, in general, the simulation agrees with experimental observation. In experiments, at low frequencies, the alignment rates first rise then fall with the increase of electric conductivity when laser intensity increases. In simulation, the maximum alignment rate and peak frequency increase with electric conductivity, corresponding to the increase of alignment rate with laser intensity in experiments. Importantly, this model also predicts the transverse alignment at ~ 750 kHz when the conductivity of silicon is $1 \times 10^{-2}$ S/m, which agrees with that observed experimentally. However, there is a discrepancy. With the increase of the laser intensity, the simulation shows that transverse alignment exhibits a monotonically increase of alignment rate and a blue shifted peak frequency. This differs from our experimental results, where the transverse alignment vanishes when the laser intensity is higher than 16 mW/cm$^2$. It could be the case that the transverse alignment frequency simply shifted to a frequency higher than 4 MHz that we test, or other mechanisms than the Maxwell-Wagner effect exist at this frequency range. It requires further investigation.

In summary, these simulation and analytic studies explain the presence of the transverse alignment observed in Si nanowires. With light induced increase of electric conductivity, the so-called photoconductivity, the overall alignment torque monotonically increases in magnitude and the spectrum shifts to higher frequencies.

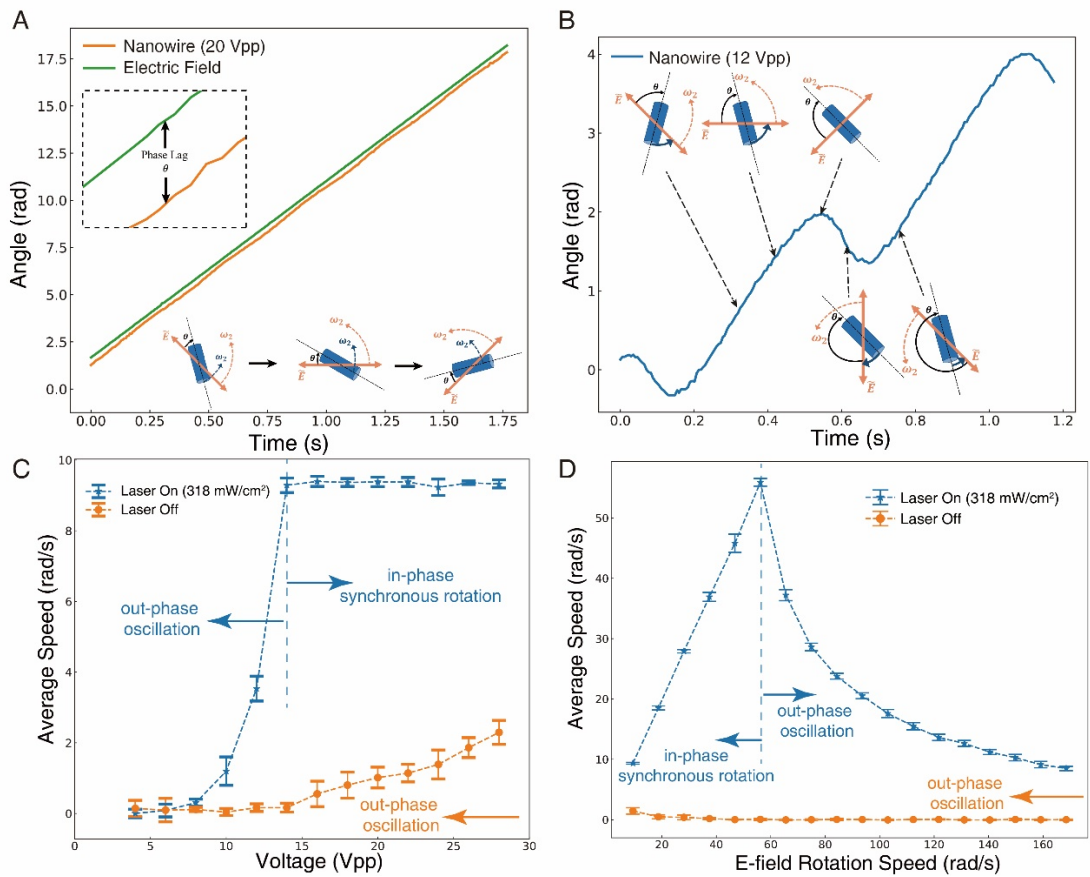

Figure 4. Light switchable in-phase and out-phase rotation of a Si nanowire stepper motor. (A,B) Rotation angle versus time of a nanowire at different voltages exposed to 318 mW/cm² 532-nm laser. (A) The electric field rotates at 1.49 Hz (in green). The stepper motor rotates synchronously with the electric field with a phase lag $\theta$, at 20 Vpp (in orange). Inset: The constant phase lag between the driving electric field and the rotating nanowire. (B) It switches to the out-phase oscillation mode at 12 V with the average speed, period and amplitude determined from the curve (SI note 5). Schematics show the relative positions of a rotating nanowire with the electric field at different time. (C) Average rotation speed of a micromotor versus driving voltage with (in blue) and without laser illumination (in orange). The driving electric field is 1.49 Hz, same as that the in-phase rotation of the nanowire. (see supplementary

video 3, 6) (D) Average rotation speed of a micromotor versus rotation speed of the driving electric field at 15 Vpp with (in blue) and without laser illumination (in orange). (see supplementary videos 4, 5)

**Light programmable synchronous stepper micromotor.**

With the understanding of the optically tunable electro-alignment of semiconductor nanowires, we propose an innovative light controlled synchronous stepper micromotor. It mimics the macroscopic synchronous stepper motors, which rotate in-phase with an electric field and turn to specific angles on demand. The operation phase of our miniaturized motors can be readily selected and switched by light. To realize such a micromotor, we generate a high frequency AC electric field ($f_1 = \frac{\omega_1}{2\pi}$) that efficiently aligns a nanowire and rotates this high frequency AC field at a much lower frequency ($f_2 = \frac{\omega_2}{2\pi} \ll f_1$) clockwise, so that the nanowire follows the low-frequency rotating AC field with a continuous synchronous rotation. By pausing the rotation of this electric field at controlled time, the motors can be stopped at any designated angle. Such an electric field can be given by:

$$\boldsymbol{E} = \mathrm{E}_0 \cos(\omega_1 t) \cdot [\cos(\omega_2 t)\,\hat{\boldsymbol{x}} + \sin(\omega_2 t)\,\hat{\boldsymbol{y}}]. \tag{5}$$

We note this driving mechanism of the motor differs from most previous works in which the rotation is a result of an electric torque generated from the interaction between the high-frequency rotating electric field and out-of-phase electric polarization of a nanowire[31,36]. Here, an in-phase electric alignment torque, is created and rotated continuously to compel the rotation[37] with the advantages of precision speed control and angular positioning, synchronism among different motors, and programmable light switching.

It is created by applying four AC voltages as shown in the following equations sequentially on the quadruple microelectrodes in Fig. 3A:

$$V_{1,3} = \pm V_0 \sin(\omega_1 t)\cos(\omega_2 t)$$
$$V_{2,4} = \pm V_0 \sin(\omega_1 t)\sin(\omega_2 t) \quad (6)$$

The voltages can be generated via a customized computer-controlled four-channel function generator (see methods and supplementary note 1). In this electric field, a nanowire keeps aligning to the high frequency component of the electric field ($\omega_1$), and continuously re-aligns to the new direction of the rotating electric field at frequency $\omega_2$ when the driving torque is sufficient. Particularly, we observe that a micromotor always operates in one of two modes, the "in-phase" and "out-phase" modes, depending on whether the alignment torque is great enough to overcome the drag. The motor runs in-phase when it rotates at the same rotating speed of $\omega_2$ ("synchronous speed") with the rotating electric field, where the alignment torque satisfies: $\tau_e = -\tau_{drag} = \gamma \omega_2$. As previously discussed, the alignment torque is a function of the angle between the nanowire and the electric field, given by: $\tau_e = -\gamma A \sin\theta \cos\theta$, so the maximum alignment torque $\tau_e$ available to drive the motor is $\frac{\gamma A}{2}$, which occurs when the angle between the nanowire long axis and the electric field is $\pi/4$. Therefore, if $\frac{\gamma A}{2} \geq \gamma \omega_2$, the electric field can supply a sufficient torque to power the motor's in-phase rotation as shown in Fig. 4A. In particular, the nanowire rotates at an approximately constant angle $\theta$ behind the electric field, where $\theta$ is termed as the "phase lag" (Fig. 4A inset). Otherwise, the alignment torque is insufficient and the nanowire, though rotates, periodically falls behind the electric field resulting in oscillations as shown in Fig. 4B. When the motor runs out-phase, it exhibits a net speed, in the same direction but lower than the rotating electric field. Although in-phase continuous rotation may seem more useful, we find many attractive features in these out-phase oscillations, which however, are not the focus of this paper. Therefore, we provide more analysis of the out-of-phase oscillation in supplementary note 5 for readers' interest.

Laser illumination control the motion of such a micromotor and allows fine control beyond basic electric manipulation. To investigate the optical effect on the rotation of the micromotor, we scan the peak-to-peak voltage from 4 Vpp to 28 Vpp for an electric field rotating at a constant speed of $\omega_2 = 1.49$ Hz. Without laser exposure, the micromotor always rotates out of phase. While its average speed increases with the voltage, it is always lower than $\omega_2$. When excited by a 318 mW/cm$^2$ 532-nm laser and the voltage is above or equal to 14 Vpp, the micromotor switches from the mode of out-phase oscillation to in-phase synchronous rotation at the same speed of the electric field ($\omega_2$) (Fig. 4C, supplementary video 3, 6).

We further scan the rotation speed of the electric field ($\omega_2$) from 1.49 Hz to 26.82 Hz at a constant voltage of 15 Vpp. Without laser, the micromotor always operates out-of-phase oscillation with nearly zero average speed. With a 318 mW/cm$^2$ 532-nm laser, the motor switches to in-phase synchronous rotation up to 8.94 Hz before losing the synchronism and turning into out-of-phase oscillation (Fig. 4D, supplementary video 4, 5). The characterizations of the performances of the stepper motors in the mode of synchronous operation, such as the pull out torque and power, are included in the supplementary note 6 for interested readers.

The two operation modes can be switched back and forth with simple light exposure without degradation in performance (see supplementary video 7). The dramatic change in the motor operation controlled by light can be leveraged to develop reconfigurable devices. For instance, with fine adjustment of the laser intensity, one could readily tune the pullout torque and power output of the motor when the motor runs in-phase. This could be achieved individually on an array of motors in the same electric field with each projected with different light patterns. A simple demonstration of individually controlled motors by light is shown as follows.

With the unique light responsiveness of the micromotors, we have successfully operated micromotors independently and versatily in the same electric field by projection of localized light spots via a digital light processing (DLP) system. The core component of the DLP system is a digital micromirror device (DMD) with an array of 1024× 768 micromirrors made by Texas Instrument (more details in supplementary note 1). The DLP system is capable of projecting micro light patterns onto the sample surface with a maximum resolution of 5 μm. To control two micromotors individually, we programmed two circular laser spots of 70 μm in diameter and each spot covers one of the micromotors without any overlap (supplementary video 9). A rotating electric field with $f_1 = 100 \text{ kHz}, f_2 = 1.49 \text{ Hz}$ at 13 Vpp is applied as the driving source. When both laser spots are on, both micromotors rotate synchronously at the same speed, as shown in Fig. 5A, where the first segment of orange and blue curves overlap, indicating the synchronous rotation speed of motor 1 and 2. Then the laser spot on motor 1 is turned off. The driving torque on motor 1 is no longer sufficient to maintain the synchronism with the electric field, and the motor 1 steps into out-phase oscillation with a lower average speed (red curve). Next, the laser spot on motor 2 is also turned off; instantly, the motor 2 is switched into out-phase oscillation (purple curve). When both laser spot 1 and 2 are turned on again, both motors restore to synchronous rotation. To the best of our knowledge, this is the first demonstration of microscale rotary stepper motors made of the same micro/nanoparticles that can be controlled on individual independently in the same environments. The success is based on modulating the electric properties of each nanoparticle device instantaneously with external stimuli, *i.e.* visible light, the interactions of which with electric fields, resulting in specific mechanical motions, *i.e.* oscillations or synchronous rotation.

Finally, such micromotors can be readily programmed and step to designated angular positions as shown in Fig. 5B and supplementary video 8. The nanowire angles are precisely controlled within five degrees. Here, the electric field (100 kHz, 10 Vpp) was directed to various angles by continuing its high-frequency component ($f_1$) but pausing the low-frequency component ($f_2$) at desired angle.

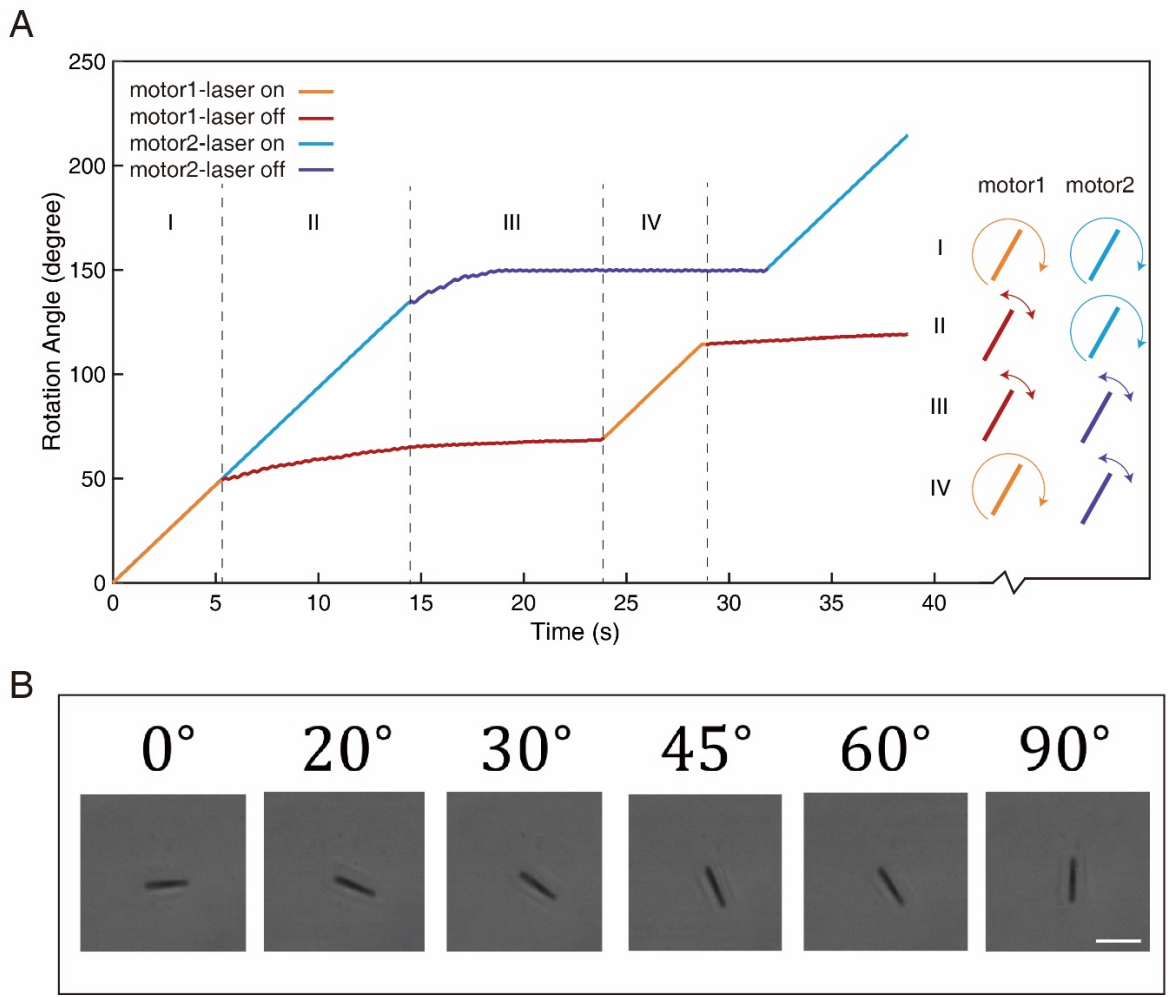

Figure 5. (A) Independent control of two stepper micromotors in the same electric field. The micromotors are both driven by a rotating electric field with $f_1 = 100$ kHz and $f_2 = 1.49$ Hz at 13 Vpp. Digital light projecting (DLP) system projects light on each micromotor independently at an intensity of ~50 mW/cm$^2$. (B) Rotation of a micromotor to designated angular positions like a stepper motor by stopping the rotation of the AC electric field at specific angle ($f_1 = 100$ kHz, V = 10 Vpp, laser intensity = 127 mW/cm$^2$), scale bar 10 μm.

## Discussion:

In summary, we report a versatile working mechanism that can be exploited for developing a new type of efficient visible-light-responsive micro/nanomachines without requirements of chemical fuels, UV light source, or special geometry of nanoparticle building blocks. The electrical property of a semiconductor micromotor can be readily controlled solely by intensity of visible light and is exhibited as modulable mechanical alignment in an external electric field. The working principle could be generally feasible for all light-responsive semiconductors. The facile switching mechanism is a new add to the tool box of nanomanipulation techniques, and could offer opportunities worth of exploration. For instance, as shown in the demonstration, encoded light signals can be transmitted into alignment oscillations of a single device, which communicates meaningful words in Morse code. The same approach could be potentially explored to develop unique opto-mechanical converters for complex coupled micro/nanomachines. Moreover, with patterned light from DLP, we have successfully demonstrated individually controlled rotary stepper motors in the same field, each can toggle between the oscillation and full-turn synchronous rotation modes, independently.  With tunable intensity of light, it is feasible to control each in the pull-out torque and power as analyzed in the supplementary note 6. Ultimately, it could be attainable to control arrays of stepper micro/nanomotors that exhibit individually specified performances by dynamic light patterns. The working mechanism could also be exploited for the future micro/nanorobots operating independently for collaborative task forces, and even to control different components within a micro/nanorobots for programmable operations.

In the aspect of fundamental research, previously, we studied the light effect on the imaginary part (out-of-phase) of the electrical polarization of silicon nanowires, exhibited by the rotational behavior in a high-frequency rotating AC electric field. However, the light effect on the real part (in phase) of the electrical polarization, which is of equal importance for many electromechanical and electrochemical phenomena, remains unknown. Thus, we conducted this study by investigating the light stimulated electro-alignment behavior of silicon nanowires, which is governed by the real-part of the electrical polarization, with both experiments and simulation. This work completes the framework of the optical effect on both real and imaginary parts of electrical polarization of semiconductor Si nanowires in aqueous solution. It is also helpful for the validation of the Kramers-Kronig relation quantitatively in the low frequency regime in the future study.

Every coin has two sides. The presented switchable manipulation is efficient and versatile, however, it is more feasible for on-chip devices than in-vivo applications. Also, a few challenges should be addressed in future research, including the oxidation problem of many semiconductors in aqueous solution, and the compatibility with complex medium[38,39]. More sophisticated design should be considered to replace the simple wires for useful applications. The full potential of this technique will also be helped by advanced optical projection devices with high resolution and large size, as well as sophisticated programming of light to equip automation[40] and even intelligence to the manipulation. Overall, this work suggests new opportunities in nanorobotics research, in both fundamental and application aspects.

Methods:

We fabricate silicon nanowires via the well-known metal-assisted chemical etching (MACE) methods as reported in our previous work[31] with a slight modification. In brief, a dispersed monolayer of 500 nm diameter polystyrene nanospheres is assembled on a cleaned undoped silicon wafer. A catalytic metal thin film of 25 nm Ag and 5 nm Au is deposited on the top such that the polystyrene nanospheres act as a mask for the following fabrication. A scotch tape is used to remove the nanospheres, leaving a metal film with circular nanoholes on the wafer. Immersing the sample into hydrofluoric acid and hydrogen peroxide dissolves the silicon underneath the metal film to leave arrays of nanowires. Finally, we use silver and gold etchants to remove the catalytic metal layer, followed by sonication in DI water to break the nanowires off the substrate. All the nanowires used in our experiments are approximately $10 \mu m$ in length, 500 nm in diameter, and composed of undoped silicon. The nanowires are stored in DI water and used in a few days after preparation. A surface oxide layer develops in a few days.

We also make two types of electrodes, "parallel" and "quadrupole", for measurement of nanowire-to-E-field alignment rate and nanowire micromotor control, respectively. The electrodes are made of 5nm Cr and 100 nm Au thin film deposited via electron beam evaporation in one of two patterns (via photolithographic mask), shown in Fig. 1D, E, on to a glass microscope slide. The parallel electrodes are two large rectangular pads separated by a long 125 um gap, while the quadrupole electrodes are four rectangular pads surrounding a square area 500 um on a side. The parallel electrodes are used for measuring nanowire-to-E-field alignment rates because the electrode configuration creates a spatially uniform electric field in the center, thereby minimizes dielectrophoretic effects, while the quadrupole electrodes are used to change the angle of the electric field to create nanowire micromotors. The parallel and quadrupole electrodes are attached via silver epoxy to an Agilent 33250A function generator (capable of 5 kHz to 4MHz signals up to 15 Vpp) and a customized computer-controlled four-output function generator (capable of arbitrary waveforms up to 2 MHz and 30 Vpp), respectively. All reported voltages assume voltage drops in the nanowire suspension between electrodes. The electric field strength is the division of the voltage by the electrode separation distance.

In each experiment, nanowires in suspension of ~20 uL is placed on top of the electrodes (either parallel or quadrupole) inside a SU8 microwell covered by a class slide. Nanowires gradually deposition to the bottom of the setup and stay in plane with the microelectrodes. A 532 nm diode-pumped solid-state laser (Thorlabs) is used for light excitation, which can be toggled off and on at different intensities. In order to reduce the influence of background light, the sample is illuminated with an always-on custom LED white light source of ~500 lx. A Basler acA1300-200um camera captures images of the electrodes and nanowires (with laser filtered out) at up to 1280x1024 pixels and up to 1000 frames per second (FPS). A computer program can track nanowire positions for data analysis via standard computer vision algorithms, and vary the voltage amplitude from the four-output function generator in real time to produce a rotating electric field. See more details in supplementary note 1.

1   Yang, G.-Z. *et al.* The grand challenges of Science Robotics. *Science Robotics* **3**, eaar7650 (2018).

2   Purcell, E. M. Life at low Reynolds number. *American journal of physics* **45**, 3-11 (1977).

3   Masliyah, J. H. & Bhattacharjee, S. *Electrokinetic and colloid transport phenomena*.  (John Wiley & Sons, 2006).

4   Ramos, A., Morgan, H., Green, N. G. & Castellanos, A. Ac electrokinetics: a review of forces in microelectrode structures. *Journal of Physics D: Applied Physics* **31**, 2338 (1998).

5   Wang, W., Chiang, T.-Y., Velegol, D. & Mallouk, T. E. Understanding the efficiency of autonomous nano-and microscale motors. *Journal of the American Chemical Society* **135**, 10557-10565 (2013).

6   Sachs, J. *et al.* Role of symmetry in driven propulsion at low Reynolds number. *Physical Review E* **98**, 063105 (2018).

7   Wang, H. & Pumera, M. Fabrication of Micro/Nanoscale Motors. *Chemical Reviews* **115**, 8704-8735, doi:10.1021/acs.chemrev.5600047 (2015).

8   Kim, K., Guo, J., Xu, X. & Fan, D. Recent Progress on Man-Made Inorganic Nanomachines. *Small* (2015).

9   Han, K. *et al.* Sequence-encoded colloidal origami and microbot assemblies from patchy magnetic cubes. *Science advances* **3**, e1701108 (2017).

10  Kim, K., Xu, X., Guo, J. & Fan, D. Ultrahigh-speed rotating nanoelectromechanical system devices assembled from nanoscale building blocks. *Nature communications* **5**, 3632 (2014).

11  Wang, B. *et al.* Reconfigurable Swarms of Ferromagnetic Colloids for Enhanced Local Hyperthermia. *Advanced Functional Materials*, 1705701 (2018).

12  Whitesides, G. M. & Grzybowski, B. Self-assembly at all scales. *Science* **295**, 2418-2421 (2002).

13  Kim, K., Zhu, F. Q. & Fan, D. Innovative mechanisms for precision assembly and actuation of arrays of nanowire oscillators. *ACS nano* **7**, 3476-3483 (2013).
28

**Acknowledgements:**

We thank S. Miao, K. Seong for the helpful technical support.

**Funding:**

We are grateful for the support of NSF via the CAREER Award (grant no. CMMI 1150767) and research grant EECS-1710922 and Welch Foundation (grant no. F-1734).

**Author contributions:**

D.F. and Z.L. conceived the research project. D.F. supervised the project. D.T. designed and implemented the customized function generator and the computer vision tracking system. Z.L designed and setup the experimental apparatus, implemented the modeling. D.T. and Z.L. conducted data analysis. All authors analyzed and discussed the results and co-wrote the manuscript.